\documentclass[5p,preprint]{elsarticle}

\usepackage{graphicx}
\usepackage{epstopdf}
\usepackage{hyperref}
\usepackage{soul}

\begin{document}
\title{Roughness-induced domain structure in perpendicular Co/Ni multilayers}
\date{\today}

\begin{abstract} We investigate the correlation between roughness, remanence
  and coercivity in Co/Ni films grown on Cu seed layers of varying thickness.
  Increasing the Cu seed layer thickness of Ta/Cu/8x[Co/Ni] thin films
  increases the roughness of the films. In-plane magnetization loops show that
  both the remanence and coercivity increase with increasing seed layer
  roughness. Polar Kerr microscopy and magnetic force microscopy reveal that
  the domain density also increases with roughness.  Finite element
  micromagnetic simulations performed on structures with periodically modulated
  surfaces provide further insight. They confirm the connection between domain
  density and roughness, and identify the microsocpic structure of the domain
  walls as the source of the increased remanence in rough films. The
  simulations predict that the character of the domain walls changes from
  Bloch-like in smooth films to N\'eel-like for rougher films.
\end{abstract}

\author[sfu]{N.~R.~Lee-Hone\corref{cor1}}
\ead{nleehone@sfu.ca}
\author[cdl]{R.Thanhoffer}
\author[ifw]{V. Neu}
\author[ifw]{R. Sch\"afer}
\author[sfu]{M. Arora}
\author[iibp]{R. H\"ubner}
\author[cdl]{D. Suess}
\author[sfu,cifar]{D. M. Broun} 
\author[sfu]{E. Girt\corref{cor2}} 
\ead{egirt@sfu.ca}

\cortext[cor1]{Corresponding author}
\cortext[cor2]{Principal corresponding author}
\address[sfu]{Department of Physics, Simon Fraser University, 8888 University
Dr, Burnaby, BC V5A 1S6, Canada}
\address[cdl]{Christian Doppler Laboratory Advanced Magnetic Sensing and MaterialsPhysics of Functional Materials  University of Vienna, 1090 Wien, W\"ahringer Stra{\ss}e 17, Austria}
\address[ifw]{IFW Dresden, Institute for Metallic Materials, Helmholtzstrasse
20, D-01069 Dresden, Germany}
\address[iibp]{Institute for Ion Beam Physics and Materials Research, Helmholtz
Zentrum Dresden Rossendorf, Dresden, Germany}
\address[cifar]{Canadian Institute for Advanced
Research, Toronto, Ontario, MG5 1Z8, Canada}

\maketitle

\section{Introduction}

Modern magnetic devices are typically multilayer structures made of a
combination of thin magnetic and non-magnetic films. Much current research
focuses on films with perpendicular magnetic anisotropy (PMA) as they provide
an avenue for creating devices that have higher storage
density~\cite{moser_magnetic_2002, piramanayagam_perpendicular_2007} and lower
switching currents~\cite{mangin_current-induced_2006} while maintaining thermal
stability. Such films have been widely studied for use in perpendicular
magnetic recording media~\cite{girt_influence_2006, richter_transition_2007,
nolan_microstructure_2007, girt_experimental_2007, neu_probing_2013}, racetrack
memory~\cite{parkin_memory_2015, parkin_magnetic_2008}, patterned magnetic
media~\cite{ross_patterned_2001}, and spin-transfer torque magnetic random
access memory (STT-MRAM) devices~\cite{khvalkovskiy_basic_2013}. Co/Ni
multilayers are one of many PMA materials that have been proposed for these
applications. Since their prediction and observation in 1992 by Daalderop et
al.~\cite{daalderop_prediction_1992}\ they have come under intense scrutiny as
they offer the possibility of low cost magnetic devices with large PMA.

Understanding the magnetic properties of thin films is complicated by the fact
that they invariably suffer from some degree of roughness when grown under
real-world conditions. It is well known that the roughness of magnetic thin
films plays a large role in determining properties such as magnetic
anisotropy~\cite{chappert_magnetic_1988, chang_jap_1994, kim_jap_1996},
coercivity~\cite{chang_jap_1994}, domain structure, magnetization reversal
process, and demagnetizing factor~\cite{chappert_magnetic_1988}. Magnetic
anisotropy has been shown to increase when films are grown on smoother
substrates~\cite{chang_jap_1994} or on smoother seed (buffer)
layers~\cite{kim_jap_1996}. The increased coercivity observed in films grown on
rougher substrates has been attributed to the pinning of domain
walls~\cite{shaw_roughness_2010}. In particular, defects at the interfaces have
been cited as the reason for this extra pinning~\cite{chang_jap_1994}. Despite
the large amount of theoretical and experimental work, the effects of roughness
are not yet fully understood. For example, there is a lack of micromagnetic
modeling on rough films, which can provide insight into domain-wall structures
that would otherwise be inaccessible to experimental techniques.

This study was undertaken to explore some of the effects that roughness has on
the magnetization reversal of thin films with PMA in the presence of an in-plane
magnetic field. We prepared PMA multilayers with well controlled roughness and
have combined this experimental study with finite element micromagnetic
simulations. Cross-sectional transmission electron microscopy (TEM) and atomic
force microscopy (AFM) were used to study the roughness; polar magneto-optical
Kerr effect (MOKE) microscopy and magnetic force microscopy (MFM) to investigate
the magnetic domain structure; and SQUID/VSM to measure the magnetization loops.
The finite element simulations confirm the experimentally observed changes in
domain size with roughness, and provide further insight into the source of
in-plane coercivity and remanence in thin films with PMA. In the following
discussion we consider three types of films: completely flat; smooth (small
roughness); and rough (large roughness).

\section{Sample preparation} The films used in this study have composition\\
Ta(3)/Cu($t_\mathrm{Cu}$)/8$\times$[Co(0.21)/Ni(0.57)]/Ta(3), with thicknesses
given in nanometers. The Cu seed layer thickness was systematically increased
as this has been shown to increase surface roughness in a controlled
manner~\cite{shaw_roughness_2010}. All layers were deposited at room
temperature on \mbox{Si (001)} wafers using both RF (Co, Ni, Ta) and DC (Cu)
magnetron sputtering at an argon pressure of $2\times10^{-3}$~torr. The base
pressure of the system was below $5\times10^{-8}$~torr. Before deposition the
substrates were cleaned with the RCA Standard Clean~1 (\mbox{SC-1})
procedure~\cite{kern_evolution_1990} to remove organic contaminants. This
leaves a thin $\rm SiO_2$ layer on the surface of the Si. The Ta and Cu seed
layers serve to establish growth along the $\left<111\right>$ directions, which
is required to obtain large PMA in Co/Ni.  In this study, the Cu seed layer
thickness, $t_\mathrm{Cu}$, was varied from 2 to 30~nm. X-ray diffraction
measurements indicate that the Co/Ni multilayers grew along the
$\left<111\right>$ directions. The films are strongly textured with the full
width at half maximum of the (111) peak below $4^{\circ}$ in an X-ray rocking
curve measurement.

\section{Results and Discussion}

\begin{figure}
  \includegraphics[width=\columnwidth]{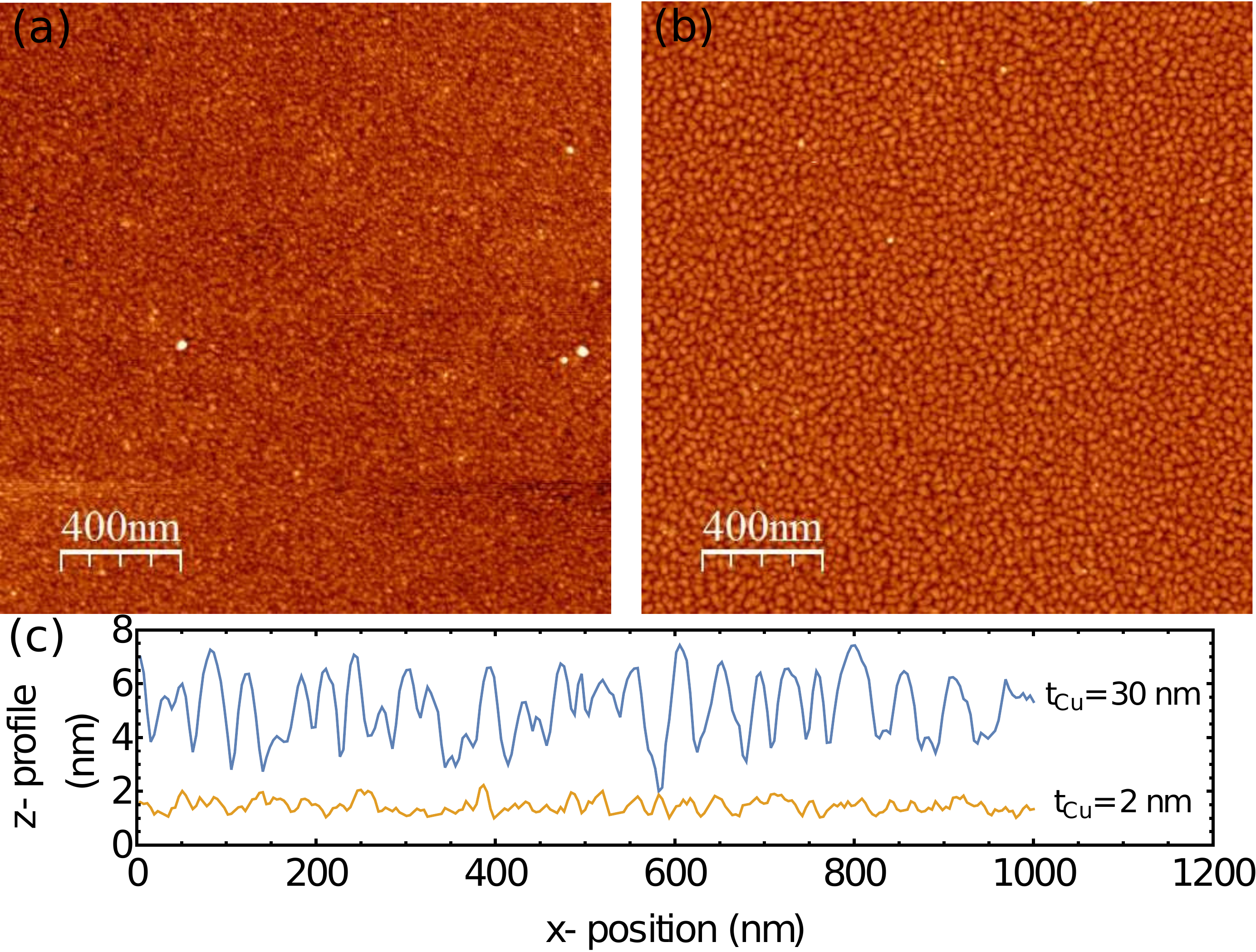}

  \caption{\label{fig:AFM}AFM images of Co/Ni multilayers grown on 2~nm (a) and
  30~nm (b) Cu seed layers. (c) Line scans through the AFM images show a large
  increase in roughness when the films are grown on thick seed layers. Profiles
  are offset for clarity.}

\end{figure}

\begin{figure}
  \includegraphics[width=\columnwidth]{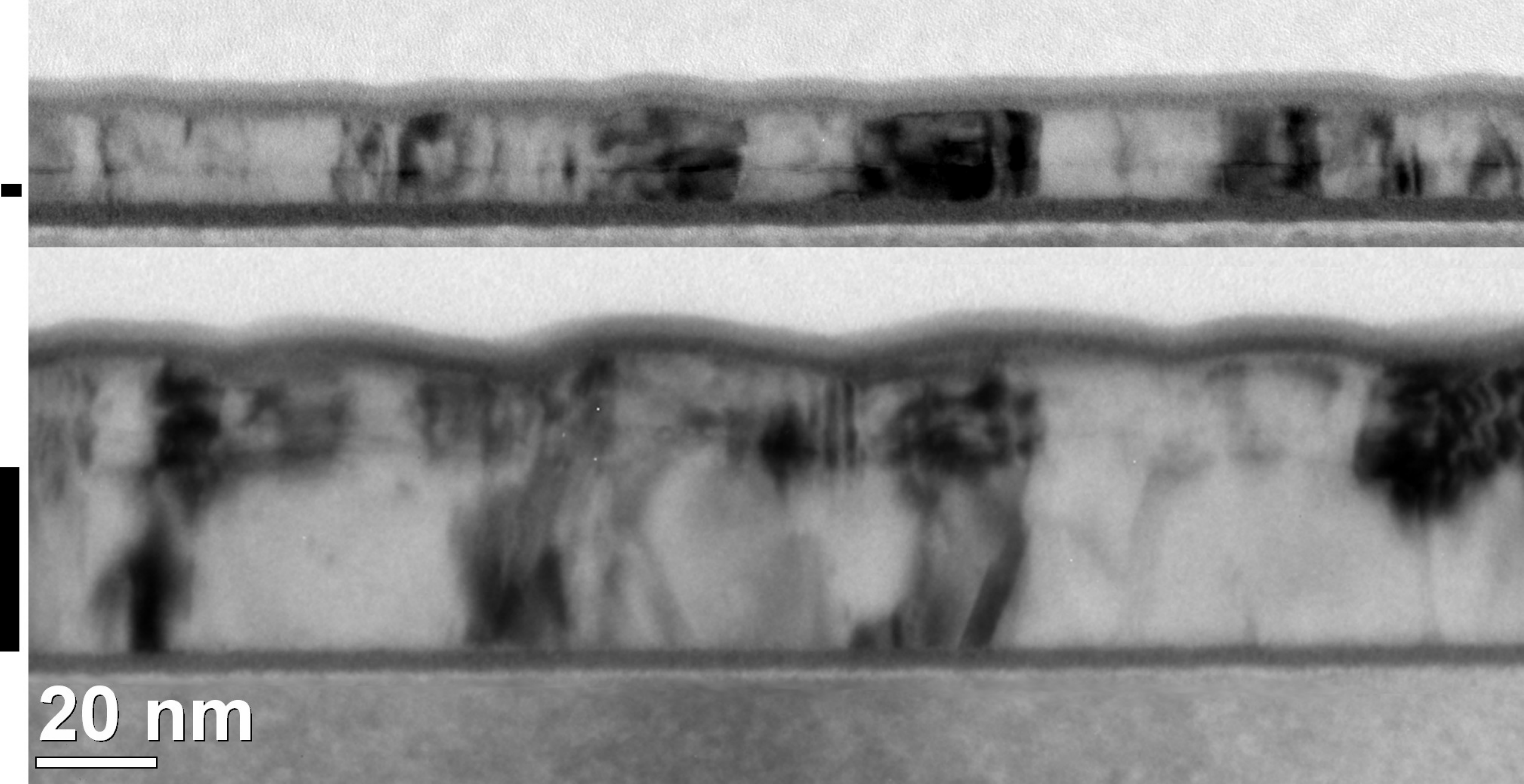}

  \caption{\label{fig:TEM}TEM images of a full Co/Ni STT-MRAM stack (see Fig.~1
  of Ref.~\cite{arora_spin_2016} for exact composition) grown on $t_\mathrm{Cu}=2$~nm
  (top) and $t_\mathrm{Cu}=30$~nm (bottom). The location and thickness of the Cu seed
  layer is indicated by the black bar on the left.}
    
\end{figure}

\autoref{fig:AFM} shows AFM images of Co/Ni multilayer films grown on 2 and
30~nm thick Cu seed layers, and corresponding topography profiles.  The RMS
roughness and grain size of the films increase with seed layer
thickness, see \autoref{tab:results}.

In addition to the films prepared specifically for this study we have included
TEM measurements from comparable Co/Ni multilayer stacks previously studied for
STT-MRAM applications that were grown on seed layers that have the same
thickness as the ones used in this study.  \autoref{fig:TEM} shows that the
Co/Ni films follow the morphology of the seed layer. The STT-MRAM films grown
on thin Cu seed layers are almost smooth with a mean grain size of 20~nm and a
mean grain height (trough to peak) of 1.6~nm. The films grown on thick Cu seed
layers are rough and have a mean grain size of 40~nm with a mean variation in
grain height of 4~nm. The modulation of the multilayer height is then similar
to the thickness of the Co/Ni multilayer.

\begin{figure*}
  \includegraphics[width=\textwidth]{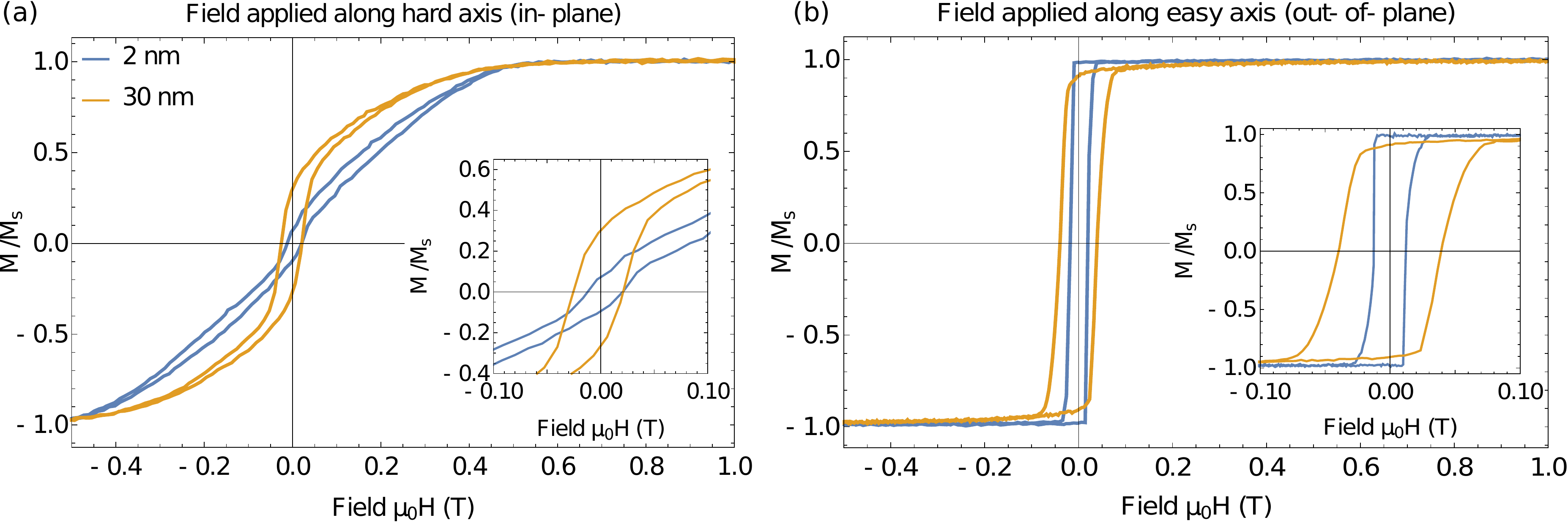}

  \caption{\label{fig:magnetization1}(a) In-plane and (b) out-of-plane
  magnetization curves measured by VSM for
  Ta(3)/Cu($t_\mathrm{Cu}$)/8$\times$[Co(0.21)/Ni(0.57)]/Ta(3) samples, grown
  on 2~nm and 30~nm Cu seed layers. Insets: zoom of the low field region.  In
  panel a, the 2~nm hysteresis loop appears to have a $\sim5$~mT exchange bias.
  This is an artefact of the measurements which were done at a sweep rate of
  4~mT/s with an averaging time of 3s.}

\end{figure*}

\begin{figure*}
  \includegraphics[width=\textwidth]{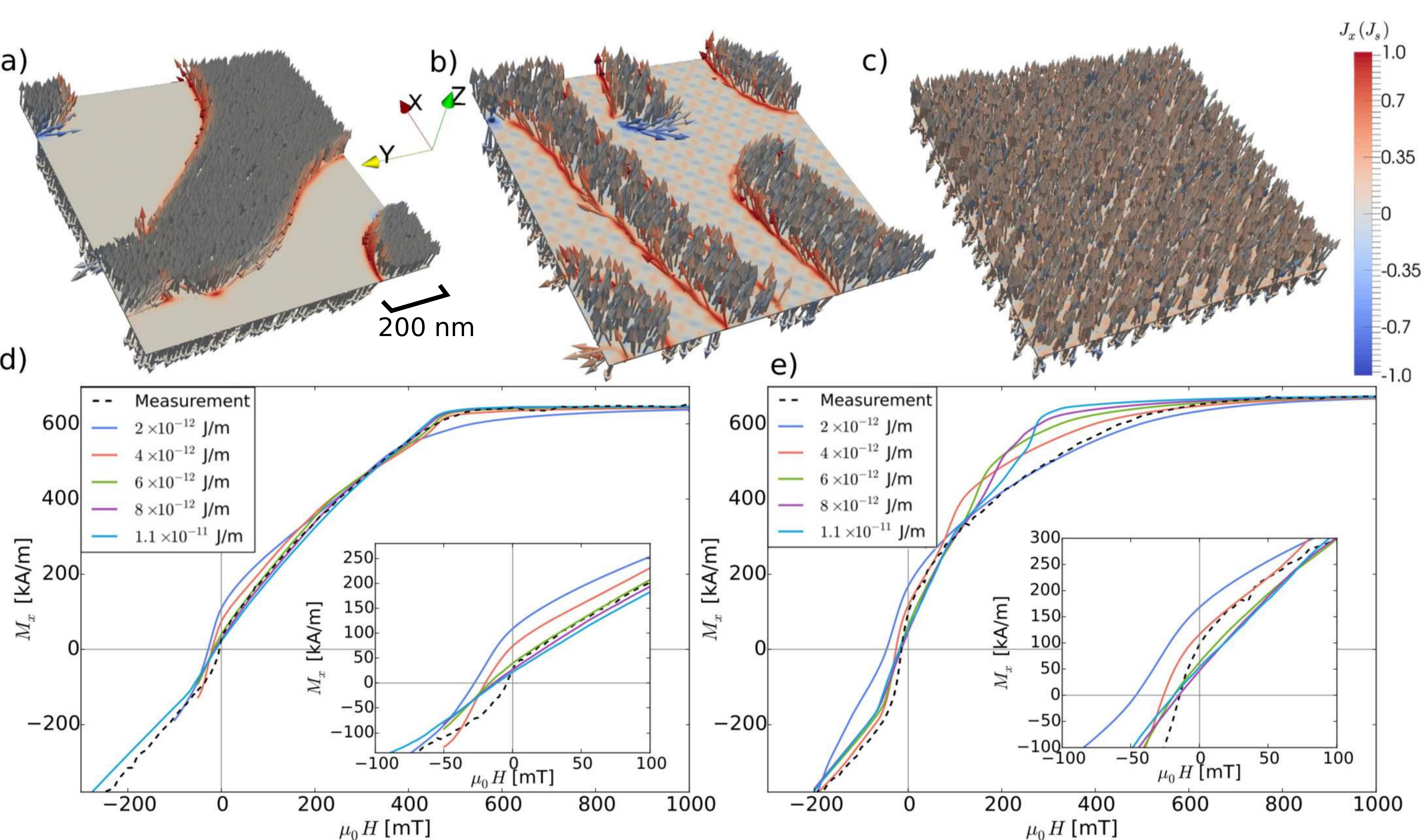}

  \caption{\label{fig:simulations} (a,b,c) Remanent magnetization state
  obtained by micromagnetic simulations after in-plane saturation. The
  $x$-component of the magnetization is color coded. In micromagnetic
  simulations with large exchange constant ($A_{\rm ex} =
  1.1\times10^{-11}$~J/m) flat (a) and rough (b) films show clear domain
  patterns. (c) By contrast, when the exchange coupling is reduced by an order
  of magnitude ($A_{\rm ex} = 2\times10^{-12}$~J/m) there is no clear domain
  pattern discernible in the simulations. Measured and simulated in-plane
  hysteresis loops with different exchange constants for (d) smooth and (e)
  rough films. Note the rounding of the simulated magnetization curves with low
  $A_{\rm ex}$.}

\end{figure*}

In-plane and out of plane hysteresis loops measured by vibrating sample
magnetometry (VSM) for the \mbox{$t_\mathrm{Cu}=2$~nm} and
$t_\mathrm{Cu}=30$~nm samples are shown in \autoref{fig:magnetization1}. The
in-plane coercivity and remanence for the rough film are more than twice as
large as those of the smooth film, and there is substantial rounding of the
hysteresis loop in the rough film. These effects are confirmed by more precise
SQUID measurements, which are presented in \autoref{fig:simulations}~(d,e). The
coercivity measured perpendicular to the film is significantly larger for the
rough film than for the smooth one, but the squareness of the hysteresis loop,
$M(0)/M_s$, is only 0.92 for the rough film as opposed to approximately 1 for
the smooth film.

\begin{table*}
  \begin{tabular}{l|cccccc}
    $t_\mathrm{Cu}$ & $M_{\rm s}$ & $M_{\rm eff}$ & $K_u$ & $\Delta H_0$ & RMS Roughness & Grain size\\
    (nm) & (kA/m) & (kA/m) & ($10^5$~J/m$^3$) & (kA/m) & (nm) & (nm) \\
    \hline
    2.0(1) & 649(13) & -337.4(2) & 4.00(5) & 7.0(4) & 0.25 & 25 \\
    12.0(3) & 655(13) & -354.4(2) & 4.15(5) & 21.5(8) & 0.73 & 37\\
    30.0(9) & 675(13) & -- & $4.17^*$ & -- & 1.17 & 43\\
  \end{tabular}

  \caption{\label{tab:results}Results of FMR and SQUID measurements for Co/Ni
  multilayers grown on top of Cu seed layers with different thickness. $\Delta
  H_0$ is the zero-frequency offset due to inhomogeneous line broadening in the
  FMR measurements as explained in Ref.~\cite{montoya_broadband_2014}.  The
  grain size is estimated from the average grain area in the AFM images.
  \mbox{*}The linewidth of the FMR data for the multilayer grown on the 30~nm
  Cu seed layer was too broad to get accurate values for $K_u$. We therefore
  calculated $K_u$ for this sample from the area enclosed between the hard axis
  and easy axis SQUID M(H) loops plus $0.5\mu_0 M_s^2$ as described in
  Ref.~\cite{johnson_magnetic_1996}.}

\end{table*}

The magnetic anisotropy of the Co/Ni films was measured using ferromagnetic
resonance (FMR) in a terminated waveguide, in a field-swept, field-modulated
configuration, as detailed by Montoya et al.~\cite{montoya_broadband_2014} The
FMR measurements were performed between 45 GHz and 69 GHz, with field applied
both parallel and perpendicular to the film surface. In order to extract the
uniaxial anisotropy, $K_u$, we first fit the frequency dependence of the
resonance and used
$$ M_{\rm eff}=M_{\rm s}-\frac{2K_u}{\mu_0M_{\rm s}} $$ where $M_{\rm eff}$
is the effective demagnetization field and $M_{\rm s}$ is the saturation
magnetization, which was measured using a SQUID magnetometer in magnetic fields
up to $\mu_0 H=5$~T. The results are shown in \autoref{tab:results}.

\begin{figure*}
  \includegraphics[width=\textwidth]{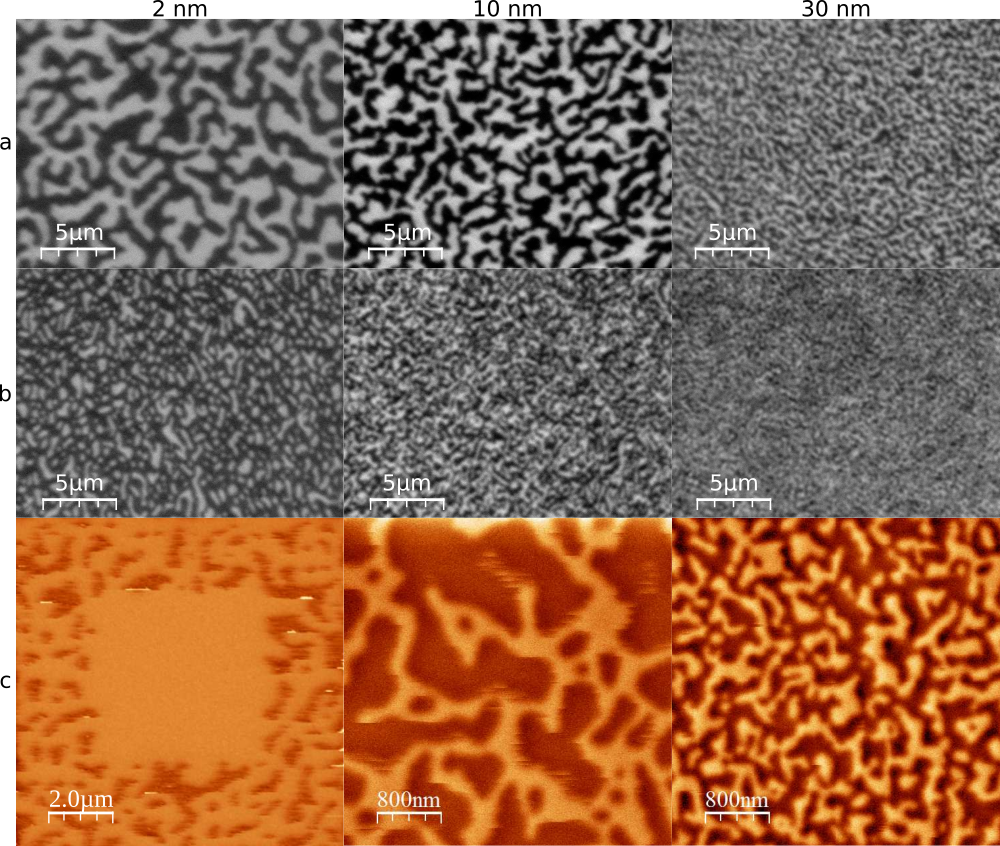}

  \caption{\label{fig:MFM}By row: (a) Polar MOKE images following out-of-plane
  ac demagnetization. (b) Polar MOKE images following in-plane saturation. (c)
  MFM images after in-plane saturation. All images were taken at zero field.
  Column headers denote the copper seed-layer thickness (2, 12 and 30~nm) of
  the respective films. The MFM images have been prepared with the help of
  WSxM, a software for scanning probe microscopy~\cite{horcas_wsxm:_2007}.}

\end{figure*}

\autoref{fig:MFM} shows the magnetic domain structure of three different films
as imaged by Kerr and magnetic force microscopy. The polar Kerr images
(\autoref{fig:MFM}~a) after ac-demagnetizing the samples in the out-of-plane
direction show the patchy up/down domain pattern often observed in thin films
with PMA, such as epitaxial L1$_0$ FePt or $c$-axis oriented $\rm SmCo_5$
films~\cite{seifert_field-_2012}. The effect of roughness shows up as a
substantial reduction in the size of the domains from about 1.2~$\mu$m average
domain size for the smoothest film ($t_\mathrm{Cu}=2$~nm) to about 400~nm for
the roughest film ($t_\mathrm{Cu}=30$~nm).  We will see later that similar
effects appear in our micromagnetic simulations.  Interestingly, also after
in-plane saturation, i.e., after a field has been applied along the hard axis,
the domain structure refines with increasing surface roughness
(\autoref{fig:MFM}~b). The average domain sizes are, however, much smaller
(from 700~nm down to 300~nm) and become difficult to resolve by optical
microscopy for the film prepared on the 30~nm thick Cu buffer. We thus resort
to imaging the same sample state with magnetic force microscopy (MFM) in a
Bruker Icon AFM/MFM with high resolution/low moment tips (225c07ML1) purchased
from Team Nanotec. The choice of tips is motivated by the small out-of-plane
coercivity of the samples, which is easily exceeded by the stray field of
standard tips when in close proximity to the film surface. Indeed, imaging the
smoothest sample failed even with the low moment tips.
\autoref{fig:MFM}~(c,~2nm) shows the MFM image of the smoothest sample after
performing a first scan on a small area and then a second scan on a larger
area; the domains were fully pushed/pulled out of the first scan area. This
means that the domain structure seen in \autoref{fig:MFM}~(c,~2nm) is not
representative of the actual domains in the film. For the slightly rougher film
($t_\mathrm{Cu}=12$~nm) the MFM measurement still displays some ‘wiping’
features corresponding to the above mentioned issue, but the out-of-plane
domain structure is already qualitatively visible. MFM performed on the
roughest sample results in a clear image of the perpendicular domains and the
small $\sim100$~nm features are clearly resolved. 

Finite element micromagnetic simulations have been used to obtain
further insights into the magnetic structure of films with PMA in in-plane
applied fields, in particular the increased domain density caused by
roughness. The lateral
dimensions of the simulated films are 1000 nm $\times$ 1000 nm, and the film
thickness is 6.24~nm to match that of the real films. To ensure that
finite element discretization was not affecting the results, we quantified the
smallest exchange lengths relevant to the simulations. Using definitions for
the exchange lengths from Abo et al.~\cite{abo_definition_2013} we took $l_{\rm
ex,ms}=\sqrt{2A_{\rm ex}/(\mu_0 M^2_{\rm s})}$ for the magnetostatic exchange
length and $l_{\rm ex,mc} = \sqrt{A_{\rm ex}/K_u}$ for the magnetocrystalline
exchange length. Here $A_{\rm ex}$ is the exchange interaction constant and
$K_u$ is the uniaxial anisotropy constant. With the experimentally determined
film properties and assuming a lower bound of $A_{\rm ex}=1\times10^{-12}$~J/m,
this gives $l_{\rm ex,ms}\sim2$~nm and $l_{\rm ex,mc}\sim1.5$~nm. Since these
values are somewhat less than the thickness of the magnetic layers in the
actual sample, there is a concern that spurious effects due to discretization
may influence the results.  To place an upper bound on these effects,
simulations were carried out for two different regimes of discretization: one
in which the magnetic film is divided vertically into three layers, and one in
which the film is treated as a single layer. No significant differences between
the 3-layer and single-layer simulations are observed, giving us confidence in
the discretization process. In particular, domain walls in the 3-layer
simulations run perpendicular to the film surface, showing that the magnetic
moments are coupled tightly together in this direction. With this consistency
check in hand, the simulations in the remainder of the paper were carried out
in the single-layer approximation, significantly reducing computation time.

Roughness has been simulated by imposing the following modulation onto the
surface: $$\Delta z=Z\sin\left(\frac{2\pi
x}{\lambda_x}\right)\sin\left(\frac{2\pi y}{\lambda_y}\right).$$ Note that this
model is a simplification of the experimental situation, in which the grains
have a distribution of heights and sizes --- this level of the simulation is
intended mainly to be illustrative of the effect of roughness.  Detailed fits
to experiment lie beyond the scope of the theory, and in any case would not be
possible because the calculations are too computationally intensive to
exhaustively explore parameter space.

The direction of uniaxial anisotropy is initially set parallel to
the local film normal and then a distortion is applied to the anisotropy
direction to simulate the effects of the grains not growing perfectly along the
$\left<111\right>$ directions. The range of distortions in the following simulations was
set to $\pm 5^{\circ}$, meaning that the local anisotropy direction deviated
from the surface normal by up to $5^{\circ}$ (with the angle chosen at random from a uniform
distribution). The simulations started by saturating the sample in the $+x$
direction in an external field of $\mu_0H_x = 1$~T. The external
field was then reduced in steps to $\mu_0H_x = -1$~T. Each simulation was run 10
times with different random distributions of the easy axis, and the hysteresis
loops in the figures are the average of the 10 runs.

\begin{figure}
  \includegraphics[width=\columnwidth]{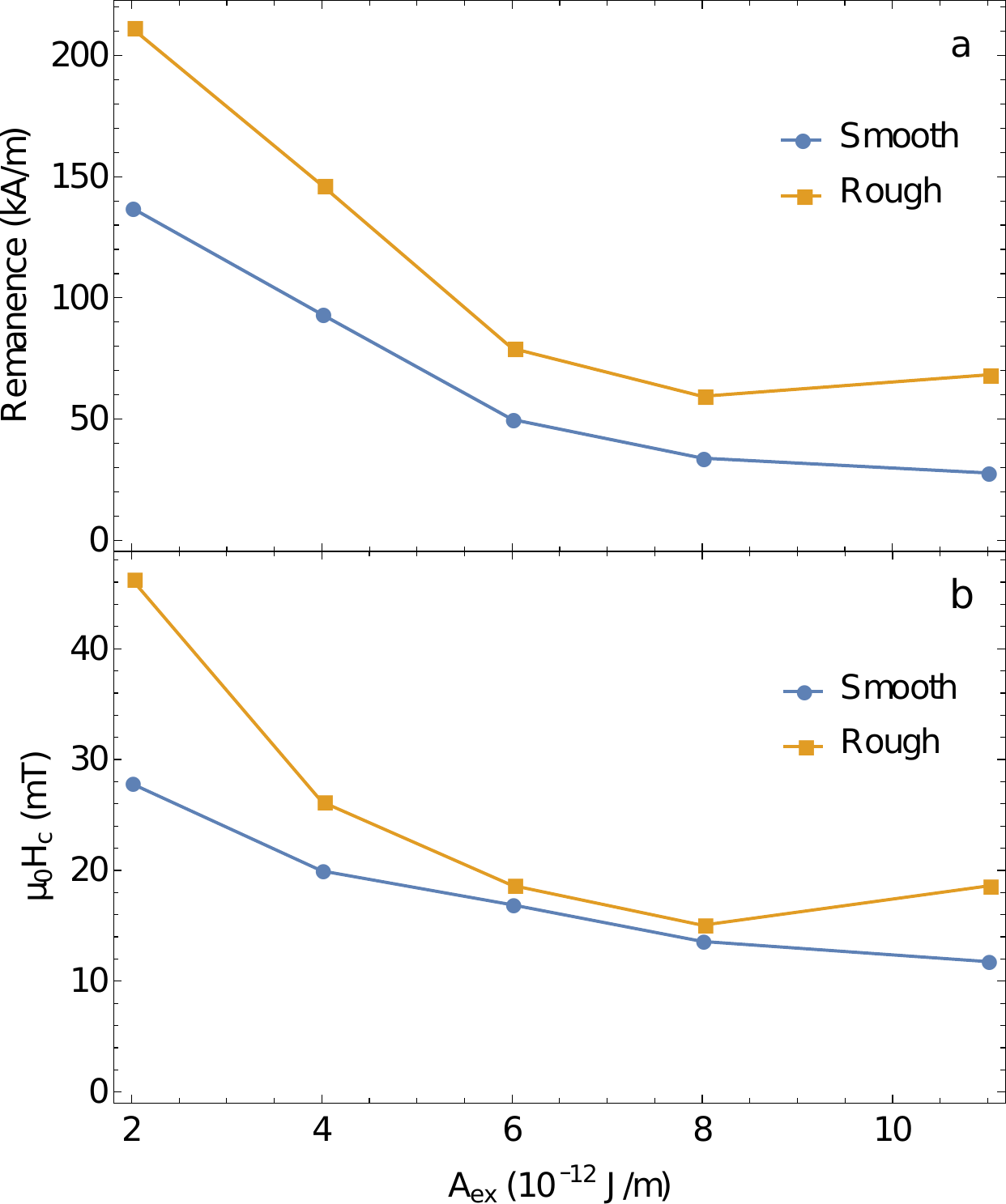}
  \caption{\label{fig:remanence_coercivity}Remanence (a) and coercivity (b)
  extracted from the simulations shown in \autoref{fig:simulations}. The rough
  films (orange squares) have higher coercivity and remanence than the smooth
  films (blue circles).} \end{figure}

\begin{figure}
  \includegraphics[width=\columnwidth]{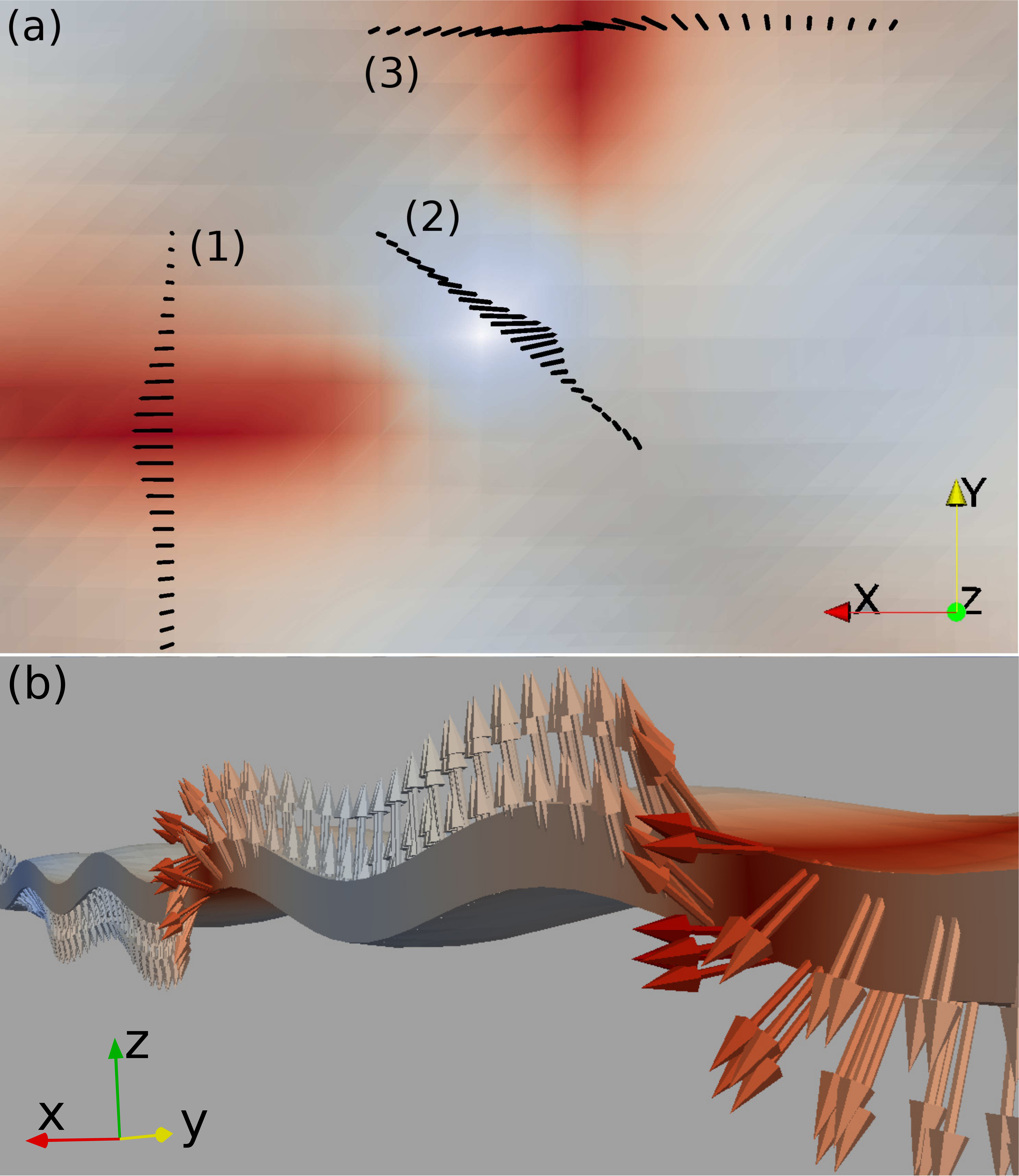}
  
  \caption{\label{fig:wall_types}(a) Top down view of three domain wall
  configurations observed in rough films after the field is removed. Red(blue)
  areas indicate a positive(negative) $x$-component of the magnetization. (1) A
  pure Bloch wall, (2) a mixed Bloch-N\'eel wall, (3) a pure N\'eel wall. (b) A
  slice through two adjacent domain walls reveals Bloch walls with reversing
  chirality.}

\end{figure}

Two sets of simulations were initially performed, one of perfectly flat
samples, the other of rough samples. The roughness parameters are $Z = 4$~nm and
\mbox{$\lambda_x=\lambda_y=80$~nm} for the rough samples. All other parameters
\mbox{($K_u$, $A_{\rm ex}$, $M_{\rm s}$)} are common. Typical domain
structures for these films in zero applied field, following saturation in the
+$x$ direction, are shown in \autoref{fig:simulations}~(a,b). The rough films
had a higher number of domains in all simulations, in agreement with the
experimental observations.

The exchange interaction determines the domain-wall width,
$\delta_w=\pi\sqrt{A_{\rm ex}/K_u}$ , and the domain-wall energy density, $E_w =
4\sqrt{A_{\rm ex}K_u}$. The competition between domain-wall energy, which
favours fewer domains, and dipole energy, which favours more domains, is
therefore tied to the exchange interaction. This is indeed observed in the
simulations: reducing the exchange coupling to $A_{\rm
  ex}=1.1\times10^{-12}$~J/m decreases the domain wall thickness, and is
accompanied by a large increase in the number of domains, almost to the point
that individual domains are indistinguishable, as shown in
\autoref{fig:simulations}~(c).

Simulations were then performed in order to understand the origin of the
nonzero remanence. To do this, we varied $A_{\rm ex}$ between
$1.1\times10^{-11}$~J/m and $0.2\times10^{-11}$~J/m, while setting $K_u$ and
$M_s$ to their experimentally determined values (see \autoref{tab:results}).
\autoref{fig:simulations}~(d,e) show that substantial rounding of the
magnetization curves is found in simulations with small exchange constant.  The
rounding of the rough film hysteresis loops therefore indicate that the
exchange constant in rough films is lower than that of the smooth films.  The
reduction of the exchange constant in the rough film can be accounted for by
the fact that the grains in the rough film have domed surfaces: this reduces
the contact area between grains, resulting in a lower effective exchange
interaction.

The larger in-plane coercivity found in the rough films is in part caused by
the reduced intergranular coupling. Roughness, however, plays a much larger
role than simply lowering $A_{\rm ex}$. In \autoref{fig:remanence_coercivity}
we see that even for the same exchange constant both the in-plane remanence and
in-plane coercivity of the rough films are larger than those of the smooth
films. It is clear that the increased coercivity and remanence cannot be
attributed to defects or impurities as the micromagnetic simulations do not
include these effects. Roughness on its own is therefore enough to increase the
coercivity and remanence.

A striking feature of all the simulations is that the magnetization in the
domain walls points predominantly in the direction of the saturating magnetic
field once the field is removed. This causes the nonzero remanence in these
films and leads to a larger remanence in the rough film since roughness
increases the number of domain walls. This observation is in accord with the
magnetization data shown in \autoref{fig:magnetization1}, where the rough film
remanence is more than twice that of the smooth film. The simulations also show
that the domain walls, which occupy only a small fraction of the total film
volume, account for approximately 75\% of the remanence, with the rest coming
from within the domains. A study of 200~nm Ni grown on 50~nm Cu by Marioni et
al.~\cite{marioni_remanence_2006} came to the same conclusions: the large
in-plane remanence is mostly due to the net magnetization of the domain walls.

In simulations of perfectly flat films the domain walls are always of Bloch
type.  This has the effect that only the component of the domain wall moments
that runs parallel to the $x$-axis (applied field direction) contributes to the
remanence.  Even in completely flat films, there is a preference for the domain
wall moments to align along the $+x$ direction, resulting in a nonzero remanence.

The situation in rough films is more complicated. In \autoref{fig:wall_types}~(a),
domain wall 1 is aligned parallel to the applied field ($x$-axis) and is of
Bloch-wall type. Domain wall 3 is aligned perpendicular to the applied field,
and is of N\'eel type. Domain wall 2, oriented at an intermediate angle, is of
mixed Bloch-N\'eel character. In all three cases there is an $x$-component of
magnetization, suggesting that all domain walls contribute to the remanence.

A consequence of the moments in the domain walls aligning predominantly in the
direction of the applied field in rough films is that the chirality of the
domain walls reverses at either side of the domain -- if it did not reverse then
the film would have no remanence. An example of the reversing chirality is shown
in \autoref{fig:wall_types}~(b). Because of the reversing chirality,
neighbouring domain walls carry opposite topological charge. The domain walls
can easily annhilate if brought close together, as opposed to the case in films
with single chirality, where at low fields topological charge can only enter and
leave the material at its
boundaries~\cite{thomas_topological_2012,kunz_field_2009}.

\section{Conclusions}
We have shown that roughness plays a crucial role in determining the in-plane
magnetic properties of thin films with PMA. Roughness increases the in-plane
remanence, coercivity, and domain density. Micromagnetic simulations link the
increased remanence to the domain density; the moments within the domain walls
point predominantly along the in-plane applied field direction, which, combined
with an increased number of domain walls, results in the increased remanence.
The simulations also predict that roughness can affect the energy balance
between N\'eel and Bloch walls. In the smooth films considered here the domains
are all Bloch-like, whereas in the rough films we also find N\'eel-like and
mixed Bloch-N\'eel walls. Further experiments are needed to verify this effect.

\section*{Acknowledgments}

The authors thank S. Pofahl for technical help with the Kerr microscope.
Financial support by the Natural Sciences and Engineering Research Council of
Canada, the Canadian Institute for Advanced Research, the Austrian Science
Fund (FWF): I2214-N20, the Austrian Federal Ministry of Science, Research and
Economy, and the National Foundation for Research, Technology and Development
is gratefully acknowledged.  Support from the structural characterization
facilities at the HZDR Ion Beam Center is gratefully acknowledged.

\bibliography{bibliography}
\bibliographystyle{unsrt}

\end{document}